\documentclass[SEDP]{cedram-sem}

\usepackage{amsfonts}
\usepackage{amsthm}
\usepackage{graphicx}
\theoremstyle{plain}
\newtheorem{theorem}{Theorem}[section]
\newtheorem{proposition}[theorem]{Proposition}

\theoremstyle{definition}
\newtheorem{Definition}[theorem]{Definition}

\newcommand{\nc}{\newcommand}

\nc{\fig}[4]
{
  \begin{figure}[ht!]  
    \centering{\scalebox{#1}{\includegraphics*{./figures/#2.eps}}}
    \caption{#4}
    \label{fig:#3}
  \end{figure}
}

\nc{\bXYZ}{{\bf XYZ\ }}

\nc{\nn}{\nonumber} 
\nc{\nit}{\noindent}
\nc{\marginnote}[1] {\marginpar{\tiny #1}}
\nc{\SH}{Schr\"odinger}
\nc{\NLS}{nonlinear Schr\"odinger}
\nc{\ie}{\emph{i.e.\ \mbox{}}}
\nc{\eg}{\emph{e.g.\ \mbox{}}}
\nc{\per}{{l_j}}
\nc{\perx}{q}
\nc{\vol}{{\rm Vol}}
\nc{\Dr}{{\rm D}}
\nc{\notDr}{{{\rm D}^c}}

\nc{\lapx}{\dfrac{\partial^2}{\partial x^2}}
\nc{\lapy}{\dfrac{\partial^2}{\partial y^2}}
\nc{\lapxy}{\dfrac{\partial^2}{\partial xy}}
\nc{\diff}[2]{\frac{d #1}{d #2}}
\nc{\diffn}[3]{\frac{d^{#3} #1}{d {#2}^{#3}}} 
\nc{\pdiff}[2]{\frac{\partial #1}{\partial #2}} 
\nc{\pdiffn}[3]{\frac{\partial^{#3} #1}{\partial{#2}^{#3}}} 

\def\Xint#1{\mathchoice
  {\XXint\displaystyle\textstyle{#1}}%
  {\XXint\textstyle\scriptstyle{#1}}%
  {\XXint\scriptstyle\scriptscriptstyle{#1}}%
  {\XXint\scriptscriptstyle\scriptscriptstyle{#1}}%
  \!\int}
\def\XXint#1#2#3{{\setbox0=\hbox{$#1{#2#3}{\int}$} \vcenter{\hbox{$#2#3$}}\kern-.5\wd0}}
\def\dashint{\Xint-}
\nc{\Av}{\dashint_\cell}
\nc{\avg}[1]{\mbox{$\left \langle\ \!#1\!\!\ \right \rangle$}}
\nc{\intRd}{\int_{{\mathbb R}^d}}

\nc{\abs}[1] {\lvert #1 \rvert} 
\nc{\norm}[2] {{\lVert #1 \rVert}_{#2}} 
\nc{\normH}[1]{\|#1\|_{H^1}^2}
\nc{\normL}[1]{\|#1\|_2^2}
\nc{\nl}[1]{|#1|^{2\sigma}}
\nc{\nlb}[1]{#1^{2\sigma+1}}
\nc{\Linvd}{L_\delta^{-1}}
\nc{\Linvz}{L_0^{-1}}
\nc{\Linvzs}{L_0^{-2}}

\nc{\st}[1]{\stackrel{(\ref{eq:#1})}{=}}
\nc{\stt}[2]{\stackrel{(\ref{eq:#1}),(\ref{eq:#2})}{=}}

\nc{\veps}{\epsilon}
\nc{\Ue}{U_\eps}
\nc{\koe}{\frac{k}{\veps}} 

\nc{\wrat}{w_{\rm ratio}}
\nc{\aij}{A^{ij}}                  
\nc{\minv}{m_*^{-1}} 
\nc{\sqminv}{m_*^{-\frac{1}{2}}}
\nc{\G}{\gamma_{\rm ef\,\!f}}
\nc{\zetap}{{\zeta_*}}
\nc{\zetas}{{\zeta_{1*}}}
\nc{\Pe}{P_{\rm edge}}
\nc{\slope}{\dfrac{d\cP[u(\cdot;\mu)] }{d\mu}}
\nc{\moot}{\mu_2}
\nc{\mum}{\mu_{\rm min}}

\def\R{\mathbb{R}}

\def\Z{\mathbb{Z}}

\nc{\brill}{{\mathcal{B}}}
\nc{\cell}{{\Omega}}
\nc{\bj}{{\bf j}}
\nc{\bk}{{\bf k}}
\nc{\bl}{{\bf l}}
\nc{\bm}{{\bf m}}
\nc{\bn}{{\bf n}}
\nc{\bkappa}{{\mathbf \kappa}}
\nc{\bK}{{\bf K}}
\nc{\bKp}{{\bf K'}}
\nc{\bp}{{\bf p}}
\nc{\bpi}{{\mathbf{\pi}}}
\nc{\bq}{{\bf q}}
\nc{\br}{{\bf r}}
\nc{\bv}{{\bf v}}
\nc{\bx}{{\bf x}}
\nc{\bX}{{\bf X}}
\nc{\bY}{{\bf Y}}
\nc{\by}{{\bf y}}
\nc{\bz}{{\bf z}}
\nc{\bnu}{{\bf \nu}}
\nc{\bxc}{{\bf x}_c}
\nc{\bxi}{{\bold\xi}}

\nc{\cP}{{\cal P}}
\nc{\cPedge}{{\cal P}_{edge}}
\nc{\cPc}{{\cal P}_{cr}} 
\nc{\cDD}{\Lambda} 
\nc{\cF}{{\cal F}}
\nc{\cG}{{\cal G}}
\nc{\cO}{{\cal O}}  
\nc{\cQ}{{\cal Q}}  
\nc{\cR}{{\cal R}}
\nc{\cI}{{\cal I}}
\nc{\cK}{{\cal K}}
\nc{\cL}{{\cal L}} 
\nc{\cM}{{\cal M}}
\nc{\cN}{{\cal N}} 
\nc{\cE}{{\cal E}}
\nc{\cH}{{\cal H}} 
\nc{\cX}{{\cal X}}
\nc{\cZ}{{\cal Z}} 
\nc{\cT}{{\cal T}} 
\nc{\order}{{\cal O}}
\nc{\tT}{{\tilde T}} 
\nc{\tlambda}{{\tilde\lambda}} 

\newtheorem{remark}{Remark}[section]

\newcommand{\x}{\mathbf{x}}

\newcommand{\eps}{\varepsilon}

\newcommand{\D}{\partial}
\newtheorem{thm}{Theorem}[section]

\nc{\pulse}{%
  \begin{picture}(60,50)(0,-20)
  \qbezier(0,0)(5,0)(10,15)
  \qbezier(10,15)(15,30)(20,30)
  \qbezier(20,30)(22,30)(25,18)
  \qbezier(25,18)(27,10)(30,10)
  \qbezier(30,10)(34,10)(38,15)
  \qbezier(38,15)(42,20)(45,20)
  \qbezier(45,20)(49,20)(53,10)
  \qbezier(53,10)(57,0)(60,0)
  \multiput(0,0)(0,-2){10}{\line(0,-1){1}}
  \multiput(60,0)(0,-2){10}{\line(0,-1){1}}
  \put(30,-20){\vector(-1,0){30}}
  \put(30,-20){\vector(1,0){30}}
  \put(0,-15){\makebox(60,10){{\tiny $T$}}}
  \end{picture}}

\title
[Waves in Honeycomb Structures]
{Waves in Honeycomb Structures}

\author
[C.L. \lastname{Fefferman}]
{\firstname{Charles} \middlename{L.} \lastname{Fefferman}}
\address{Department of Mathematics\\
Princeton University\\
Princeton, NJ 08540\\
USA}
\thanks{CLF is partially supported by US-NSF Grant DMS-09-01040.}
\email{cf@math.princeton.edu}
\author
[M.I. \lastname{Weinstein}]
{\firstname{Michael} \middlename{I.} \lastname{Weinstein}}
\address{Department of Applied Physics\\ and Applied Mathematics\\
Columbia University\\
New York, NY 10027\\
USA}
\thanks{MIW is partially supported by US-NSF Grant DMS-10-08855.}
\email{miw2103@columbia.edu}
\keywords{Periodic structure, Dispersion relation, Dirac point, Dirac equations, Conical point, Graphene}
  
\subjclass{00X99}

\begin{document}


\begin{abstract}
  We review recent work of the authors on the non-relativistic Schr\"odinger equation with a  honeycomb lattice potential, $V$. In particular, we summarize results on (i) the existence of Dirac points, conical singularities in dispersion surfaces of $H_V=-\Delta+V$
   and (ii) the two-dimensional Dirac equations, as a large, but finite time,  effective description of $e^{-iH_Vt}\psi_0$, for data $\psi_0$, which is spectrally localized at a Dirac point.
  We conclude with a formal derivation and  discussion of the effective large time evolution  for the nonlinear Schr\"odinger - Gross Pitaevskii equation  for small amplitude initial conditions, $\psi_0$. The   effective dynamics are governed by a nonlinear Dirac system. \end{abstract}

\maketitle


\section{Introduction}

There has been intense interest within the fundamental and applied  physics communities in the propagation of waves in honeycomb structures.  Two areas where such structures have been explored extensively are (i) condensed matter physics and (ii) photonics:
\begin{enumerate}
\item Graphene, a single atomic layer of carbon atoms, is a two-dimensional structure with carbon atoms located at the sites of a honeycomb structure with remarkable electronic properties; see, for example, the survey articles \cite{RMP-Graphene:09,Novoselov:11}.
\item Photonic / electro-magnetic propagation (linear and nonlinear) has been studied in dielectric  honeycomb structures \cite{Haldane:08,Segev-etal:07,Segev-etal:08,Rechtsman-etal:12}.
\end{enumerate}

The basic mathematical model is a wave equation, defined on the 2-dimensional plane with a medium whose material properties vary periodically according to a honeycomb pattern: Schr\"odinger's equation with honeycomb lattice potential, in the quantum setting, and Maxwell's equations, with honeycomb structured dielectric parameters, in the electromagnetic setting. \medskip

We focus on the time-evolution of the Schr\"odinger equation 
 \begin{equation}
 i\D_t\psi = \left(-\Delta +V(\bx)\right)\psi\ =\ H_V\psi \ ,
 \label{td-schroedinger}
 \end{equation}
 where $V(\bx)$ is a  smooth, real-valued and periodic potential, defined on $\R^2$. Denote by $\Omega$,
 a choice of elementary period cell of $V(\bx)$.
We are interested in properties of the time-evolution, which relate
 directly to the special symmetry of the honeycomb lattice.
  To represent the time evolution operator, $e^{-i H_V t}$, we quickly review the basic Floquet-Bloch spectral theory of periodic potentials.  \medskip
 
\begin{figure}[ht!]
\centering
 \includegraphics[width=4.55in]{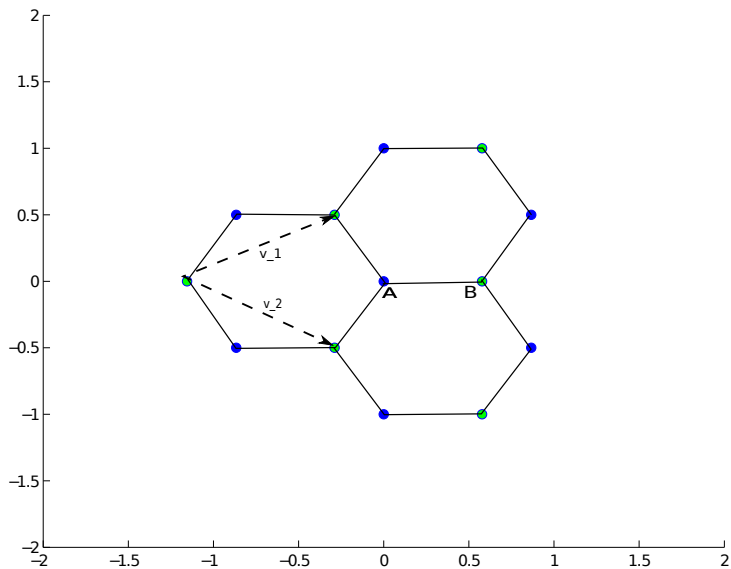}
  \caption{ Part of the honeycomb structure, ${\bf H}$.
  ${\bf H}$ is the union of two sub-lattices  $\Lambda_{\bf A}={\bf A}+\Lambda_h$ (blue) 
   and $\Lambda_{\bf B}={\bf B}+\Lambda_h$ (green). The lattice vectors 
    $\{\bv_1,\bv_2\}$ generate $\Lambda_h$. 
   } .  \label{fig:honeyAB-1}
\end{figure}

\begin{figure}[ht!]
\centering 
\includegraphics[width=4.55in]{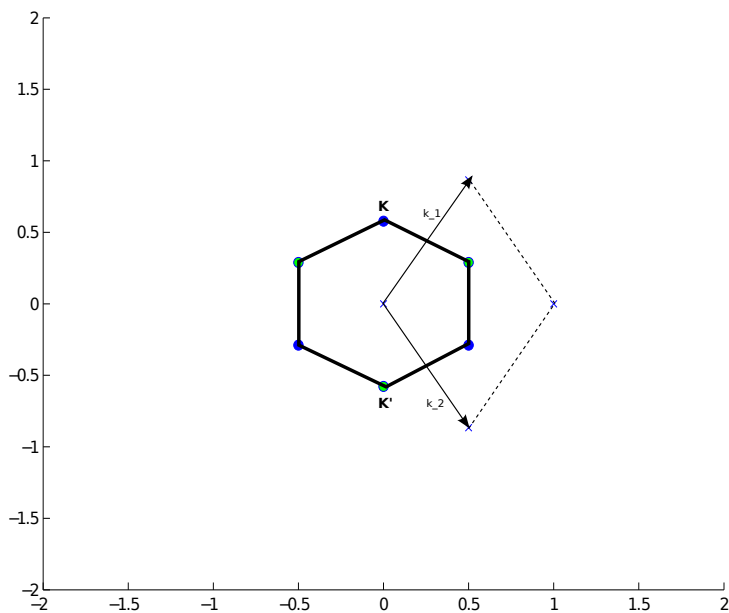}
  \caption{  Brillouin zone, $\brill$, and dual basis $\{\bk_1,\bk_2\}$. $\bK$ and $\bK'$ are labeled. Other vertices of $\brill_h$ obtained via application of $R$, rotation by $2\pi/3$.}  \label{fig:honeyAB-2}
\end{figure}

Consider the family of $\bk-$ pseudo-periodic eigenvalue problems:
\begin{align}
H_V\  \Phi(\bx;\bk) &= \mu(\bk)\ \Phi(\bx;\bk),\ \ \bx\in\R^2,\label{phi-eqn}\\
 \Phi(\x+\bv;\bk) &= e^{i\bk\cdot \bv}\ \Phi(\bx;\bk),\ \ \bv\in \Lambda.   \label{pseudo-per}
 \end{align}
 We introduce the family of $\bk-$ pseudo-periodic $L^2$ functions, $L^2_\bk$:
 \[L^2_\bk\ =\ \left\{ f\in L^2(\Omega)\ :\ f(\bx+\bv)=e^{i\bk\cdot\bv} f(\bx),\ \bv\in\Lambda \right\}\ .\]
Here,
 \begin{enumerate}
 \item $\Lambda$ denotes the period lattice for $V(\bx)$. We denote by 
 \[\Lambda_h\ =\ \Z\bv_1\oplus\Z\bv_2\ ,\ \textrm{where}\ \ \bv_1=\left( \begin{array}{c} \frac{\sqrt{3}}{2} \\ {}\\  \frac{1}{2}\end{array} \right),\ \ \bv_2 =\ \left(\begin{array}{c} \frac{\sqrt{3}}{2} \\ {}\\ -\frac{1}{2} \end{array}\right)  \]
is  the period lattice of the regular honeycomb structure; see Figure \ref{fig:honeyAB-1},  and
 \item $\bk-$ varies of the Brillouin zone, $\brill$, a choice of fundamental cell in the dual period lattice, $\Lambda^*$. For honeycomb lattice potentials, 
\[\Lambda^*_h\ =\ \Z\bk_1\oplus\Z\bk_2\ ,\ \textrm{where}\ \ {\bf k}_1=\ \frac{4\pi}{\sqrt{3}}\left(\begin{array}{c} \frac{1}{2}\\ {}\\ \frac{\sqrt{3}}{2}\end{array}\right),\ \  {\bf k}_2 = \frac{4\pi}{\sqrt{3}}\ \left(\begin{array}{c}\frac{1}{2}\\ {}\\ -\frac{\sqrt{3}}{2}\end{array}\right),\]
and  $\brill$ can be chosen to be a regular hexagon, centered at the origin. The vertices of $\brill$ fall into two equivalence class: points of $\bK$ type and points of $\bK'$ type, which alternate around the perimeter of $\brill$;
 see Figure \ref{fig:honeyAB-2}. 
 \end{enumerate}
 An equivalent spectral problem with periodic boundary conditions may be obtained by setting $\Phi(\bx;\bk)=e^{i\bk\cdot\bx}\ p(\bx;\bk)$.  Then,  $p(\bx;\bk),\ \bk\in\brill$ is periodic and satisfies
  boundary value problem: 
\begin{align}
H_V(\bk) p(\bx;\bk)\ &=\   \mu(\bk)\ p(\bx;\bk),\ \  \bx\in\R^2,\label{p-eqn}\\
  p(\bx+\bv;\bk) &= p(\bx;\bk),\ \ \bv\in \Lambda \label{p-periodic}, 
\end{align}
where
\begin{equation}
H_V(\bk)\equiv -\left(\nabla+ i\bk\right)^2 + V(\bx)\ .
\label{HVdef}
\end{equation}

For each $\bk\in\brill$, the spectrum of $H_V(\bk)$ is discrete, consisting of real eigenvalues:
\[ \mu_1(\bk)\le\mu_2(\bk)\le\cdots\le\mu_b(\bk)\le\cdots\]
(listed with multiplicities), with corresponding eigenfunctions $p_b(\bx;\bk),\ b\ge1$.
As $\bk$ varies over $\brill$, $\mu_b(\bk)$ sweeps out a closed real interval , $\mu_b(\brill)$. The spectrum of $H_V$ is the union of these intervals:
\begin{equation}
{\rm spectrum}\left(H_V\right)\ =\ \bigcup_{b\ge1} E_b(\brill)
\label{spec-HV}\end{equation}

 The Floquet-Bloch modes
$\Phi_b(\bx;\bk)=e^{i\bk\cdot\bx}p_b(\bx;\bk),\ b\ge1$ are complete in the sense that for any $f\in L^2(\R^2)$
\begin{equation}
 f(\bx)\ -\ \sum_{1\le b\le N}\ \int_{\mathcal{B}}\  
\left\langle \Phi_b(\cdot;\bk),f(\cdot)\right\rangle_{L^2(\R^2)}\ \Phi_b(\bx;\bk)\ d\bk\ \to\ 0
\label{spectraldecomp}
\end{equation}
in $L^2(\R^2)$ as $N\uparrow\infty$. The solution of the initial value problem for time-dependent Schr\"odinger equation,  \eqref{td-schroedinger}, with initial data $\psi(\bx,0)=\psi_0(\bx)$,  has the representation :
\begin{equation}
e^{-iHt}\ \psi_0\ =\ \sum_{b\ge1}\ \int_{\mathcal{B}}\ e^{-i\mu_b(\bk) t}\ 
\left\langle \Phi_b(\cdot;\bk),\psi_0(\cdot)\right\rangle_{L^2(\R^2)}\ \Phi_b(\bx;\bk)\ d\bk\ .
\label{propagator1}
\end{equation}

 .

\noindent An understanding of the time-dynamics \eqref{propagator1} requires, in particular, a detailed understanding of the functions $\mu_b(\bk),\ b\ge1$. These are called {\it band dispersion functions} of $H_V$.  The graphs $\bk\mapsto \mu_b(\bk)$ are called the {\it dispersion surfaces} of $H_V$.  \medskip

 Let's now introduce the class of potentials that interests us. For any function $f$, defined on $\mathbb{R}^2$,  introduce
\begin{equation}
\mathcal{R}[f](\bx)=f(R^*\bx),\label{calRdef}
\end{equation}
where $R$ is the $2\times2$ rotation matrix, which clockwise-rotates a vector by an angle of $2\pi/3$.
\medskip

 \begin{Definition}[Honeycomb lattice potentials]\label{honeyV}
Let $V$ be  real-valued and  $V\in C^\infty(\R^2)$.
$V$ is  a  {\it  honeycomb lattice potential} 
 if there exists $\bx_0\in\mathbb{R}^2$ such that $\tilde{V}(\bx)=V(\bx-\bx_0)$ has the following properties:
\begin{enumerate}
\item $\tilde{V}$ is $\Lambda_h-$ periodic, {\it i.e.}  $\tilde{V}(\bx+\bv)=\tilde{V}(\bx)$ for all $\bx\in\mathbb{R}^2$ and $\bv\in\Lambda_h$.  
\item $\tilde{V}$ is even or inversion-symmetric, {\it i.e.} $\tilde{V}(-\bx)=\tilde{V}(\bx)$.
\item  $\tilde{V}$  is $\mathcal{R}$- invariant, {\it i.e.}
 \begin{equation}
 \mathcal{R}[\tilde{V}](\bx)\ \equiv\ \tilde{V}(R^*\bx)\ =\ \tilde{V}(\bx),
 \nn\end{equation}
  where, $R^*$ is the counter-clockwise rotation matrix by $2\pi/3$, {\it i.e.} $R^*=R^{-1}$.
 \end{enumerate} 
 Thus, a honeycomb lattice potential is smooth, $\Lambda_h$- periodic and, with respect to some origin of coordinates, both inversion symmetric and $\mathcal{R}$- invariant.
 \end{Definition}

In this article we summarize recent results of the authors on the spectral properties of $H_V$ and properties of the evolution operator, $e^{-iH_Vt}$, where $V$ denotes a honeycomb lattice potential.  In section \ref{sec:dirac-pts} we describe results on the existence of conical singularities  in the dispersion surfaces of honeycomb lattice potentials for quasi-momenta located at the vertices of $\brill$. In section \ref{sec:dirac-eqn} we consider the initial value problem for the non-relativistic Schr\"odinger equation, \eqref{td-schroedinger}, with a honeycomb lattice potential. We study the evolution for  initial conditions which are spectrally localized about a vertex ($\bK$- type or $\bK'-$ type point) of $\brill$ and show that the large, but finite, time-evolution is effectively governed by the constant coefficient two-dimensional Dirac equation. 
 Finally in section \ref{sec:nonlinear-dirac} we remark on some open problems concerning the relation
  of the nonlinear Schr\"odinger / Gross-Pitaevskii equation and the two-dimensional nonlinear Dirac equation.
  \medskip
  
  \nit{\bf Acknowledgements:}\ The authors wish to thank  M. Ablowitz, A.C. Newell and G. Uhlmann for stimulating discussions. \medskip

 \section{Dirac points}\label{sec:dirac-pts}
 
 In this section we discuss results of the authors in \cite{FW:12} on the conical singularities, so-called {\it Dirac points}, in the dispersion surfaces of $H_V$, where $V$ is a honeycomb lattice potential. 
 
 A key property of honeycomb lattice potentials, $V$ is that if $\bK_\star$ denotes any vertex of $\brill$, then we have the commutation relation:
  \begin{equation}
  \left[\mathcal{R},H_{V}(\bK_\star)\right]\ =\ 0\ .
  \label{commutation}
  \end{equation}
  Since $\mathcal{R}$ has eigenvalues $1,\tau$ and $\overline{\tau}$, it is natural to the split $L^2_{\bK_\star}$, the space of $\bK_\star$- pseudo-periodic functions, into the direct sum:
  \begin{equation}
   L^2_{\bK_\star}\ =\ L^2_{\bK_\star,1}\oplus L^2_{\bK_\star,\tau}\oplus L^2_{\bK_\star,\overline\tau},\label{L2-directsum}
   \end{equation}
   where $L^2_{\bK_\star,\sigma}$ are invariant eigen-subspaces of $\mathcal{R}$, {\it i.e.}
  for $\sigma=1,\tau,\overline{\tau}$, where $\tau=\exp(2\pi i/3)$, and
   \begin{equation}
   L^2_{\bK_\star,\sigma}\ =\ \Big\{g\in  L^2_{\bK_\star}: \mathcal{R}g=\sigma g\Big\}\ .
\label{L2Ksigma}   \end{equation}
  
    We next give  a precise definition of a Dirac point. 
\begin{Definition}\label{Diracpt-def}
Let $V(\bx)$ be a smooth,  real-valued, even (inversion symmetric) and  periodic potential on $\R^2$.
Denote by $\brill$, the Brillouin zone.
We call $\bK\in\brill$ a {\it Dirac point} if the following holds:
There exist an integer $b_1\ge1$, a real number $\mu_\star$, and strictly positive  numbers, $\lambda$ and $\delta$, such that:
\begin{enumerate}
\item $\mu_\star$ is a degenerate eigenvalue of $H$ with $\bK-$ pseudo-periodic boundary conditions.
\item $\textrm{dim Nullspace}\Big(H-\mu_\star I\Big)\ =\ 2$
\item $\textrm{Nullspace}\Big(H-\mu_\star I\Big)\ =\ 
{\rm span}\Big\{ \Phi_1(\bx) , \Phi_2(\bx)\Big\}$, where $\Phi_1\in L^2_{\bK,\tau}$ and $\Phi_2(\bx)=\overline{\Phi_1(-\bx)}\in L^2_{\bK,\bar\tau}$. 
\item There exist Lipschitz functions $\mu_\pm(\bk)$, \[ \mu_{b_1}(\bk)=\mu_-(\bk)\ \ \  
\mu_{b_1+1}(\bk)=\mu_+(\bk),\ \ \mu_\pm(\bK)=\mu_\star\]
 and $E_\pm(\bk)$, defined for $|\bk-\bK|<\delta$, and 
  $\bk-$ pseudo-periodic eigenfunctions of $H$: $\Phi_\pm(\bx;\bk)$, with corresponding eigenvalues $\mu_\pm(\bk)$ such that
  \begin{align}
\mu_+(\bk)-\mu(\bK)\ &=\ +\ \lambda\ 
\left| \bk-\bK \right|\ 
\left(\ 1\ +\ E_+(\bk)\ \right)\ \ {\rm and}\nn\\
\mu_-(\bk)-\mu(\bK)\ &=\ -\ \lambda\ 
\left| \bk-\bK \right|\ 
\left(\ 1\ +\ E_-(\bk)\ \right),\label{cones}
\end{align}
where $|E_\pm(\bk)|\le C|\bk-\bK|$ for some $C>0$. 
  \end{enumerate}
\end{Definition}
\medskip

\begin{remark}\label{strategy-diracpt}
In \cite{FW:12} we prove the following
\begin{proposition}\label{lambda-is=|lambdasharp|}
Suppose conditions $1. , 2.$ and $3.$ of Definition \ref{Diracpt-def} hold and denote by 
$\{c(\bm)\}_{\bm\in\mathcal S}$ the sequence of $L^2_{\bK,\tau}$ Fourier-coefficients of $\Phi_1(\bx)$. 
Define the sum
\begin{equation}
\lambda_\sharp\ \equiv\   \sum_{\bm\in\mathcal{S}} c(\bm)^2\ \left(\begin{array}{c}1\\ i\end{array}\right)\cdot \bK_\star^\bm
\label{lambda-sharp}
\end{equation}
If $\lambda_\sharp\ne0$, then 
$4.$ of Definition \ref{Diracpt-def} holds (see \eqref{cones} )  with $\lambda=|\lambda_\sharp|$.\\
\end{proposition}
\end{remark}

Therefore Dirac points are found by verifying conditions $1. , 2.$ and $3.$ of Definition \ref{Diracpt-def} and the additional (non-degeneracy) condition:  $\lambda_\sharp\ne0$. We use this characterization to prove  the following result, Theorem 5.1 of \cite{FW:12}, concerning the existence of Dirac points for Schr\"odinger operators with a generic  honeycomb lattice potentials:

\begin{thm}\label{main-thm} 
Let $V(\bx)$ honeycomb lattice potential. Assume further that the Fourier coefficient of $V$, $V_{1,1}$, is non-vanishing, {\it i.e.}
\begin{equation}
V_{1,1}\ \equiv\ \int_\Omega e^{-i(k_1+k_2)\cdot\by}\ V(\by)\ d\by\ \ne0\ .
\label{V11eq0}
\end{equation}
Consider the one-parameter family of honeycomb Schr\"odinger operators defined by:
\begin{equation}
H^{(\epsilon)}\ \equiv\ -\Delta + \epsilon\ V(\bx)\ .
\label{Heps-def}
\end{equation}
There exists a countable and 
closed set $\tilde{\mathcal{C}}\subset\mathbb{R}$ such that for    all $\epsilon\notin\tilde{\mathcal{C}}$, the vertices, $\bK_\star$, of $\brill_h$ are Dirac points in the sense of Definition \ref{Diracpt-def}.\medskip

More specifically,  the following holds for $\epsilon\notin\tilde{\mathcal{C}}$:
 There exists $b_1\ge1$ such that $\mu_\star\equiv\mu^\epsilon_{b_1}(\bK_\star)=\mu^\epsilon_{b_1+1}(\bK_\star)$ is a $\bK_\star-$ pseudo-periodic eigenvalue of multiplicity two
  where
\begin{enumerate}
\item   
\subitem $\mu^\epsilon_\star$  is an $L^2_{\bK,\tau}$ - eigenvalue of $H^{(\epsilon)}$ of multiplicity one, and corresponding eigenfunction, 
$\Phi^\epsilon_1(\bx)$.
\subitem $\mu^\epsilon_\star$ is an $L^2_{\bK,\bar\tau}$ - eigenvalue of $H^{(\epsilon)}$ of multiplicity one, with corresponding eigenfunction, $\Phi^\epsilon_2(\bx)=\overline{\Phi^\epsilon_1(-\bx)}$.
\subitem $\mu^\epsilon_\star$ is \underline{not} an $L^2_{\bK,1}$- eigenvalue of $H^{(\epsilon)}$.
\item There exist $\delta_\epsilon>0,\ C_\epsilon>0$
  and Floquet-Bloch eigenpairs: $(\Phi_+^\epsilon(\bx;\bk), \mu_+^\epsilon(\bk))$  
 and $(\Phi_-^\epsilon(\bx;\bk), \mu_-^\epsilon(\bk))$, and  Lipschitz continuous functions, $E_\pm(\bk)$,   defined for  $|\bk-\bK_\star|<\delta_\epsilon$,  such that 
\begin{align}
\mu^\epsilon_+(\bk)-\mu^\epsilon(\bK_\star)\ &=\ +\ |\lambda^\epsilon_\sharp|\ 
\left| \bk-\bK_\star \right|\ 
\left(\ 1\ +\ E^\epsilon_+(\bk)\ \right)\ \ {\rm and}\nn\\
\mu^\epsilon_+(\bk)-\mu^\epsilon(\bK_\star)\ &=\ -\ |\lambda^\epsilon_\sharp|\ 
\left| \bk-\bK_\star \right|\ 
\left(\ 1\ +\ E^\epsilon_-(\bk)\ \right),\nn
\end{align}
where 
\begin{equation}
\lambda_\sharp^\epsilon\ \equiv\   \sum_{\bm\in\mathcal{S}} c(\bm,\mu^\epsilon,\epsilon)^2\ \left(\begin{array}{c}1\\ i\end{array}\right)\cdot \bK_\star^\bm\ \ne\ 0
\label{lambda-sharp2}
\end{equation}
 is given in terms of $\{c(\bm,\mu^\epsilon,\epsilon)\}$, the $L^2_{\bK,\tau}$- Fourier coefficients of $\Phi^\epsilon(\bx;\bK_\star)$.

Furthermore, $|E_\pm^\epsilon(\bk)| \le C_\epsilon |\bk-\bK_\star|$. 
Thus, in a neighborhood of the point $(\bk,\mu)=(\bK_\star,\mu_\star^\epsilon)\in \R^3 $, the dispersion surface is closely approximated by a  circular {\it cone}.
\item 
There exists $\epsilon^0>0$, such that for all $\epsilon\in(-\epsilon^0,\epsilon^0)\setminus\{0\}$\\
  (i)\  $\epsilon V_{1,1}>0\ \implies$ conical intersection of $1^{st}$ and $2^{nd}$ dispersion surfaces \\
 (ii)\ $\epsilon V_{1,1}<0\ \implies$ conical intersection of $2^{nd}$ and $3^{rd}$ dispersion surfaces\ .
 \end{enumerate}
\end{thm}

\begin{remark}\label{deformed-dirac-pt}
In a forthcoming article, we  present  a general analytic perturbation theory of deformed honeycomb lattice Hamiltonians, for perturbations which commute with inversion composed with complex conjugation. Conical (Dirac) points persist for small  perturbations of this type, although the conical singularities typically perturb away from the vertices of $\brill$. These results extend those of \cite{FW:12} and, in particular, include the case of a uniformly strained honeycomb structure. 
\end{remark}

\nit{\it Remarks on the proof of Theorem \ref{main-thm} \cite{FW:12}:}\   Following the strategy outlined in Remark \ref{strategy-diracpt}, we first show, for all $\epsilon$ sufficiently small and non-zero, by a Lyapunov-Schmidt reduction, that the degenerate, multiplicity three eigenvalue of the Laplacian with $\bK_\star-$ pseudo-periodic boundary condition, splits into a multiplicity two eigenvalue and a multiplicity one eigenvalue, with  associated $\mathcal{R}$- invariant eigenspaces. Next we must show the persistence of this degenerate multiplicity two subspace as $\epsilon$ is increased without bound.
  To  continue  to $\eps$ large we introduce a globally-defined analytic function, $\mathcal{E}(\mu,\epsilon)$, whose zeros, counting multiplicity, are the eigenvalues of $H^{(\epsilon)}$.  Eigenvalues occur where an operator 
  $I+ \mathcal{C}(\mu,\epsilon),\ \ \mathcal{C}(\mu,\epsilon)$  compact, is singular. 
 Since $\mathcal{C}(\mu,\epsilon)$ is not trace-class but is Hilbert-Schmidt, we work with $\mathcal{E}(\mu,\epsilon)=
 \det_2(I+\mathcal{C}(\mu,\epsilon))$, a renormalized determinant. 
Next $\mathcal{E}(\mu,\epsilon)$ and $\lambda_\sharp^\epsilon$  are studied using techniques of complex function theory to establish the existence of Dirac points for arbitrary real values of $\epsilon$, except possibly for  a countable closed subset, $\mathcal{C}\subset\mathbb{R}$. The origin of this exceptional set is a  topological obstruction to the existence of continuous choice of eigenvector as parameters are varied. 
\medskip

Previous analyses of such honeycomb lattice structures are based upon extreme limit models:
 the tight-binding / infinite contrast limit (see, for example, \cite{Wallace:47,RMP-Graphene:09,Kuchment-Post:07}) in which the potential is  taken to be concentrated at lattice points or edges of a graph; in this limit, the dispersion relation has an explicit analytical expression, or 
 the weak-potential limit, treated by formal perturbation theory in
\cite{Haldane:08,Ablowitz-Zhu:11} and rigorously in \cite{Grushin:09}.

 \section{Evolution of wave packets and the two-dimensional Dirac equation}\label{sec:dirac-eqn}
 
From the previous section, we see that generic honeycomb Hamiltonians, $H_V$, to have  Dirac points,
and that some band dispersion function is {\it approximately} that of a two-dimensional wave equation
 with  wave speeds $\pm|\lambda_\sharp|$. Thus, initial conditions which are spectrally localized about a Dirac point, should evolve {\it approximately} according to a constant coefficient hyperbolic PDE with 
 wave speeds $\pm|\lambda_\sharp|$. In this section we summarize results 
 deriving the appropriate wave equation and its time-scale of validity in terms of the degree of spectral localization, $\delta$, about a Dirac point, $\bK_\star$.
 
A general solution of the time-dependent Schr\"odinger equation, constrained to the  degenerate $2$-dimensional eigenspace associated with eigenvalue, $\mu_\star=\mu(\bK_\star)$, associated with the Dirac point,  $\bK_\star$,   is of the form
\begin{equation}
\psi(\bx,t)=e^{-i\mu_\star t}\ \left(\ \alpha_1\ \Phi_1(\bx) + \alpha_2\ \Phi_2(\bx)\ \right),
 \label{espace-evolve}
 \end{equation}
 where $\alpha_1$ and $\alpha_2$ are arbitrary constants.

Consider now a  {\it wave packet} initial condition, which is spectrally concentrated near $\bK_\star$:
\begin{align}
\psi^\delta(\bx,0)\ =\ \psi_0^\delta(\bx)\ &=\ \delta\ \left(\ \alpha_{10}(\delta\bx)\ \Phi_1(\bx) + \alpha_{20}(\delta\bx)\ \Phi_2(\bx)\ \right)\nn\\
&=\delta\ \left(\ \alpha_{10}(\delta\bx)\ p_1(\bx) + \alpha_{20}(\delta\bx)\ p_2(\bx)\ \right)\ e^{i\bK_\star\cdot\bx}\
 \label{packet-data}
 \end{align}
 Here, $\delta$ is a small parameter. We assume $\alpha_{10}(\bX)$ and $\alpha_{20}(\bX)$ are Schwartz functions of $\bX$. We expect that this assumption can be weakend considerably without difficulty. The overall factor of $\delta$ in \eqref{packet-data}  is not essential (the problem is linear), but is inserted so that $\psi^\delta_0$
  has $L^2(\R^2)$- norm of order of magnitude one.

We seek solutions of \eqref{td-schroedinger}, \eqref{packet-data} in the form:
\begin{equation}
\psi^\delta(\bx,t)=e^{-i\mu_\star t}\left( \sum_{j=1}^2\ \delta\ \alpha_j(\delta\bx,\delta t)\Phi_j(\bx) + \eta^\delta(\bx,t)\right).
 \label{packet-ansatz}
 \end{equation}
\begin{equation}
\textrm{where}\ \eta^\delta(\bx,0)=0,\ \alpha_j(\bX,0)=\alpha_{j0}(\bX),\ j=1,2 
\label{eta0}
\end{equation} 
to ensure the initial condition \eqref{packet-data}. 
\medskip

The goal is to show 
that the Schr\"odinger equation \eqref{td-schroedinger} has a solution of the form \eqref{packet-ansatz} with an error term, $\eta^\delta(\bx,t)$, which satisfies
\begin{equation}
\sup_{0\le t\le \rho\delta^{-2+\eps_1} }\ \|\ \eta^\delta(\cdot,t)\ \|_{H^s(\R^2)}\ =\ 
\mathcal{O}(\delta^{\tau_\star}),\ \ \delta\to0\ .
\label{eta-goal}
\end{equation}
for some $\tau_\star>0$, 
  provided  the slowly varying amplitudes $\alpha_j(\delta\bx,\delta t),\ j=1,2$ evolve according to the system of Dirac-type equations \eqref{Dirac-1}-\eqref{Dirac-2}.   Here, $\rho>0$ and $\eps_1>0$ are arbitrary.
\medskip

 In \cite{FW-dirac-validity} we prove
 
 \begin{theorem}\label{effective-Dirac} Assume 
 \[ \vec{\alpha}_0(\bX)\ \equiv\ \left(\begin{array}{c}\alpha_{10}(\bX)\\ 
 \alpha_{20}(\bX)\end{array}\right) \in \left[\mathcal{S}(\R^2)\right]^2\]
 and let $\vec{\alpha}(\bX,T)$ denote the global-in-time solution of the Dirac system
 \begin{align}
 \D_T\alpha_1(\bX,T)\ &=\ -\overline{\lambda_\sharp}\ \left(\D_{X_1}+i\D_{X_2}\right)\ \alpha_2(\bX,T)\label{Dirac-1}\\ 
\D_T\alpha_2(\bX,T)\ &=\ -\lambda_\sharp\ \left(\D_{X_1}-i\D_{X_2}\right)\ \alpha_1(\bX,T)\  ,
\label{Dirac-2}
\end{align}  
where $0\ne\lambda_\sharp\in\mathbb{C}$. %
 with initial data $\vec\alpha(\bX,0)=\vec\alpha_0(\bX)$. 
 
 Consider the time-dependent Schr\"odinger equation, \eqref{td-schroedinger}, where $V(\bx)$ denotes a potential for which the conclusions of Theorem \ref{main-thm} hold, {\it e.g.} 
 $V(\bx)=\epsilon V_h(\bx)$, where $V_h$ is a honeycomb lattice potential satisfying $V_{1,1}\ne0$ and  $\epsilon$ is not in the countable closed set $\tilde{\mathcal{C}}$. 
 
  Assume initial conditions, $\psi_0$,  of the form
  \eqref{packet-data}. 
 Fix any $\rho>0, \eps_1>0$ and $s\ge1$.  
 Then, \eqref{td-schroedinger} has a unique solution of the form \eqref{packet-ansatz}, where for any $|\alpha|\le N$
 \begin{equation}
 \sup_{ 0\le t\le \rho\ \delta^{-2+\eps_1} } 
 \left\|\ 
  \eta^\delta(\cdot,t)\ \right\|_{H^s(\R^2)}\  \ =\ o(\delta^{\tau_\star}),\ \ \delta\to0\,
  \label{eta-estimate}
 \end{equation}
 for some $\tau_*>0$. 
 \end{theorem}
 
 \begin{remark}\label{deformed-evolution}
 In connection with Remark \ref{deformed-dirac-pt}, 
we  also consider the analogous question of the dynamics of solutions for wave-packet initial data, spectrally concentrated at a Dirac point of the deformed honeycomb structure. In this case, the  large, but finite, time dynamics are being give by {\it tilted-} Dirac equations.  The latter can be mapped to the standard 2D Dirac equations by a Galilean change of variables. 
 \end{remark}
 
{\it Remarks on the proof of Theorem \ref{effective-Dirac}:}\ We seek a solution of the initial value problem with wave-packet initial condition \eqref{packet-data} of the form \eqref{packet-ansatz}, a slow space/time modulation of the degenerate subspace plus an error term $\eta^\delta(x,t)$. 
The Dirac equations \eqref{Dirac-1}-\eqref{Dirac-2} arise as a non-resonance condition, which ensures that $\eta^\delta(\bx,t)$ is small on a time interval: $0\le t\le\mathcal{O}(\delta^{-2+})$. Estimation of $\eta^\delta$,
 via its DuHamel integral equation,  requires a careful decomposition of the propagator, $e^{-i(-\Delta+V)t}$ and analysis of its action on functions with quasi-momentum components supported near $\bK_\star$, a vertex of $\brill$, and those with quasi-momentum components supported away from $\bK_\star$. The resonant terms, which are removed by imposing equations \eqref{Dirac-1}-\eqref{Dirac-2}, arise from quasi-momenta near $\bK_\star$. A detailed expansion of the normalized Floquet-Bloch modes for such quasi-momenta is required. Such modes are discontinuous at  $\bK_\star$. Components corresponding to quasi-momenta away from $\bK_\star$ are controlled, via Poisson summation and integration by parts with respect to time, by making use of rapid phase oscillations in time. \medskip 

Formal derivations of Dirac-type dynamics for honeycomb lattice structures are discussed in the physics   \cite{RMP-Graphene:09} and  applied mathematics \cite{Ablowitz-Zhu:11} literature. A rigorous discussion of the tight-binding limit is presented in \cite{Ablowitz-Curtis-Zhu:12}. 
Conical singularities have long been known to occur in Maxwell equations with constant anisotropic dielectric tensor;
  see, for example, \cite{Indik-Newell:06},  \cite{Berry-Jeffrey:07} and references cited therein. 
  
 \section{NLS / GP and the nonlinear Dirac equation}\label{sec:nonlinear-dirac}
 
 A model of fundamental importance in the description of macroscopic quantum phenomena  and nonlinear optical phenomena
  is the {\it nonlinear 
  Schr\"odinger / Gross-Pitaevskii equation}, (NLS/GP): 
 
\begin{equation}
i\D_t\psi\ =\ \left(-\Delta +V(\bx)\right)\psi+\ g|\psi|^2\psi
\label{nls-honey1}\end{equation}
We refer to the case $V\equiv0$ as the nonlinear Schr\"odinger equation (NLS).
In the quantum setting, the potential, $V(\bx)$, models a magnetic trap confining 
 a large number of quantum particles. A wide range of potentials, so called optical lattices,  can be induced optically via interference of optical beams. We consider the case where $V=V_h$, a honeycomb lattice potential in the sense of Definition \ref{honeyV}.
  Analogously, in the setting of nonlinear electromagnetics interference of beams in photorefractive  crystals can generate
a spatially dependent  index of refraction, corresponding to a honeycomb lattice potential. See, for example, \cite{PS:03,ESY:07,Newell-Maloney,Segev-etal:08}.
\medskip

Our goal in this section is to give a formal derivation of a nonlinear  Dirac equation from the NLS/GP
and to pose some interesting questions to consider going forward.
%
%

As earlier, we assume wave-packet initial conditions for NLS/GP, with amplitude scaled
so that nonlinear effects and the Dirac dynamics enter on the same time-scale:
\begin{align}
\psi^\delta(\bx,t)\ &=\ \left(\ \delta^{1\over2}\sum_{j=1}^2\alpha_j(\bX,T)\Phi_j(\bx)\ +\ \eta^\delta(\bx,t)\ \right)e^{-i\mu_\star t}\ ,\label{ansatz}
\end{align}
where $\bX = \delta\bx$ and $T=\delta t$. 
Substitution of the Ansatz \eqref{ansatz} into NLS/GP, \eqref{nls-honey1}, yields $\eta^\delta=\delta^{1\over2}\left(\delta\eta_1+\mathcal{O}(\delta^2)\right)$, where  $\eta_1$, satisfies:
\medskip 

\begin{align}
&i\D_t\ \eta_1(\bx,t)-(H-\mu_\star)\ \eta_1(\bx,t)\nn\\
&=\ -\sum_{j=1}^2\left(\  i\D_T\alpha_j\ \Phi_j(\bx) +2\nabla_\bX\alpha_j\cdot\nabla_\bx\Phi_j(\bx)\  -\ g\sum_{1\le i,l\le2}\ \alpha_i\alpha_j\overline{\alpha_l}\ \Phi_i(\bx)\Phi_j(\bx)\overline{\Phi_l(\bx)}\ \right)\ .
\label{nl-eta-eqn}\end{align}

A calculation which is analogous to that carried out and rigorously justified in \cite{FW-dirac-validity} 
for the linear Schr\"odinger equation (see  section \ref{sec:dirac-eqn})
yields the following semilinear Dirac system governing the slowly modulating amplitudes $\alpha_1(\bX,T)$ and $\alpha_2(\bX,T)$:
\begin{align}
 \D_T\alpha_1\ &=\ -\overline{\lambda_\sharp}\ \left(\D_{X_1}+i\D_{X_2}\right)\ \alpha_2
 \ -\ i\ g\left( \beta_1|\alpha_1|^2+2\beta_2|\alpha_2|^2\right)\alpha_1\ ,
 \label{nldA}\\ 
\D_T\alpha_2\ &=\ -\lambda_\sharp\ \left(\D_{X_1}-i\D_{X_2}\right)\ \alpha_1\ -
\ i\ g\left( 2\beta_2|\alpha_1|^2 + \beta_1|\alpha_2|^2\right)\alpha_2\ .
\label{nldB}
\end{align} 
The coefficients $\beta_1$ and $\beta_2$ are given in terms of the Floquet-Bloch modes:
\begin{equation}
\beta_1=\int_\Omega |\Phi_1(\bx)|^4\ d\bx\ =\  \int_\Omega |\Phi_2(\bx)|^4\ d\bx\ \ {\rm and}\ \ 
\beta_2=\int_\Omega |\Phi_1(\bx)|^2|\Phi_2(\bx)|^2\ d\bx\ .\label{beta12}
\end{equation}
\medskip

The system \eqref{nldA}-\eqref{nldB} has the following conserved integrals:
\begin{align}
&\mathcal{N}\left[\alpha_1,\alpha_2\right]\ \equiv\ \int_{\R^2}\ \left(\ \left|\alpha_1\right|^2\ +\ \left|\alpha_1\right|^2\ \right)\ dX_1\ dX_2\ , \label{L2}\\
&\mathcal{H}\left[\alpha_1,\alpha_2\right]\equiv\ {\rm Im}\left[\ \overline{\lambda_\sharp}\int_{\R^2}\alpha_2\ \left(\D_{X_1}+i\D_{X_2}\right)\overline{\alpha_1}\ dX_1\ dX_2\ \right] \ \nn\\
&\qquad\qquad \ -\ \frac{g}{4}\ \int_{\R^2}\ \left(\ \beta_1\ |\alpha_1|^4 \ +\ 4\beta_2 |\alpha_1|^2\  |\alpha_2|^2
 \ +\ \beta_1\ \left|\alpha_2\right|^4\ \right)\ dX_1\ dX_2\ .
\label{Energy}
\end{align}

A word about the calculation leading to \eqref{nldA}-\eqref{nldB}: Heuristically speaking,  the modulation equations \eqref{nldA}-\eqref{nldB} for $\alpha_j(\bX,T),\ j=1,2$, are obtained by choosing their slow, $(\bX,T)$, evolution to cancel all resonant components of the forcing term in \eqref{nl-eta-eqn}.
This can be implemented by
setting the projection of the source term onto the modes $\Phi_j,\ j=1,2$ (integrating only over the fast variable, $\bx$) to zero.

In computing this projection, many cubic terms drop by symmetry considerations.
 For example, consider the contribution of the cubic term:
$\alpha_1\alpha_1\overline{\alpha_2}\Phi_1\Phi_1\overline{\Phi_2}$ in \eqref{nl-eta-eqn} to the equation for $\alpha_1$. To compute this contribution take the $L^2(\R^2_\bx)$ inner product with $\Phi_1(\bx)$:
\begin{align}
\left\langle \Phi_1,\Phi_1\Phi_1\overline{\Phi_2}\right\rangle_{L^2(\Omega)}\ &=\ 
 \int_\Omega \overline{\Phi_1(\bx)}\cdot \Phi_1(\bx) \Phi_1(\bx) \overline{\Phi_2}(\bx)\ d\bx\nn\\
&=^{\bx=R^*\by} \int_{R\Omega} \overline{\Phi_1(R^*\by)}\cdot \Phi_1(R^*\by) \Phi_1(R^*\by) \overline{\Phi_2(R^*\by)}\ d\by\nn\\
&=\ \int_{R\Omega} \overline{\tau\Phi_1(\by)}\cdot \tau\Phi_1(\by)\cdot \tau\Phi_1(\by)\cdot \overline{\bar\tau\Phi_2(\by)}\ d\by\nn\\
&=\ \tau^2\ \int_\Omega \overline{\Phi_1(\bx)}\cdot \Phi_1(\bx) \Phi_1(\bx) \overline{\Phi_2}(\bx)\ d\bx\ =\ 
 \tau^2\ \left\langle \Phi_1,\Phi_1\Phi_1\overline{\Phi_2}\right\rangle_{L^2(\Omega)}\nn
 \end{align}
 Therefore, 
 \begin{align}
&(1-\tau^2)\left\langle \Phi_1,\Phi_1\Phi_1\overline{\Phi_2}\right\rangle_{L^2(\Omega)}\ = 0\ \implies\  \ \left\langle \Phi_1,\Phi_1\Phi_1\overline{\Phi_2}\right\rangle_{L^2(\Omega)}=0.
\nn\end{align}
It follows that a term of the form $\alpha_1 \overline{\alpha_2} \alpha_1$ does not appear in the \eqref{nldA}.\medskip

 An alternative formal approach to obtaining \eqref{nldA}-\eqref{nldB} is by the method of multiple scales; see  \cite{Ablowitz-Zhu:11} where a formal derivation of a semilinear Dirac system
for a small amplitude (``shallow'') honeycomb lattice potential of a special form is presented and studied numerically.
\medskip

As in the proof of Theorem \ref{effective-Dirac}, a proof of the large time validity of \eqref{nldA}-\eqref{nldB} 
requires that $\eta^\delta$ remains small in a suitable norm on a large (growing as $\delta\downarrow0$)  time scale. This  rests on a study of the time-evolution of the  Floquet-Bloch components of $\eta^\delta$ which are  near $\bK_\star$ and away from $\bK_\star$, together with an understanding of bounds on the solutions to the nonlinear Dirac equations \eqref{nldA}-\eqref{nldB}.
\medskip

\bigskip

We conclude with some natural questions. 
 Is there global well-posedness for the initial value problem for semi-linear Dirac system \eqref{nldA}-\eqref{nldB} or do solutions develop singularities (blow up) in finite time? How do the corresponding solutions of NLS / GP behave?
 
%
\bibliographystyle{plain}

\bibliography{fw-biarretz}
\end{document}